\documentclass{iau}
\usepackage{graphicx}

\title[Evolution of magnetic fields in solar-type stars] 
{The evolution of surface magnetic fields in young solar-type stars}

\author[Folsom et al. ]   
{C.P. Folsom$^{1}$, P. Petit$^{1}$, J. Bouvier$^{2}$, J.-F. Donati$^{1}$, J. Morin$^{3}$}

\affiliation{
$^{1}$Institut de Recherche en Astrophysique et Plan\'etologie, Toulouse, France\\
email: {\tt colin.folsom@irap.omp.eu} \\
$^{2}$Institut de Plan\'etologie et d'Astrophysique de Grenoble, France\\
$^{3}$Institute of Astrophysics, Georg-August-University, G\"ottingen, Germany\\
}

\pubyear{2013}
\volume{302}  
\pagerange{1--2}
\jname{Magnetic fields throughout stellar evolution}
\editors{P. Petit ... eds.}
\begin{document}

\maketitle

\begin{abstract}

The surface rotation rates of young solar-type stars decrease rapidly with age from the end of the pre-main sequence though the early main sequence.  This suggests that there is also an important change in the dynamos operating in these stars, which should be observable in their surface magnetic fields.  Here we present early results in a study aimed at observing the evolution of these magnetic fields through this critical time period.  We are observing stars in open clusters and stellar associations to provide precise ages, and using Zeeman Doppler Imaging to characterize the complex magnetic fields.  Presented here are results for six stars, three in the in the $\beta$ Pic association ($\sim$10 Myr old) and three in the AB Dor association ($\sim$100 Myr old). 

\end{abstract}

\firstsection 
\section{Introduction}

Solar-type stars undergo a dramatic evolution in their rotation rates as they leave the pre-main sequence and settle into their main sequence lives.  Since these stars have dynamo driven magnetic fields, there is likely an important evolution in their magnetic properties over the same time period.  In turn, stellar magnetic fields play a key role in angular momentum loss.  Thus to fully understand the angular momentum evolution of these stars, it is critical to characterize their magnetic evolution.  No study has yet traced the evolution of these magnetic properties in solar-type stars from the pre-main sequence through the early main sequence with any precision.  

In order to investigate this, we are obtaining a series spectropolarimetric observations of solar type stars in open clusters using the ESPaDOnS instrument at the Canada France Hawaii Telescope.  Focusing on stars in open clusters allows us to place precise ages on the targets.  Using a time series of observations and the Zeeman Doppler Imaging (ZDI; \cite[Donati \& Brown 1997]{Donati1997}; \cite[Donati et al. 2006]{Donati2006}) technique we can characterize the complex magnetic field geometries of the targets.  Here we present early results from this project, focusing on on 3 stars in the $\beta$ Pic association ($\sim$10 Myr old; \cite[Torres et al. 2008]{Torres2008}) and 3 stars in the AB Dor association ($\sim$100 Myr old; \cite[Torres et al. 2008]{Torres2008}; \cite[Barenfeld et al. 2013]{Barenfeld2013}).  

\section{Analysis}

Least Squares Deconvolution (LSD) was applied to our observations, and the resulting `mean' line profiles were used to measure longitudinal magnetic fields and radial velocities. The longitudinal magnetic fields were used to determine a rotational period for each star, which was checked against the phasing of the LSD profiles, and the radial velocity variability.  Stellar parameters: $T_{\rm eff}$, $\log g$, microturbulence, and $v \sin i$, were derived by fitting synthetic spectra computed with the ZEEMAN code to the observed spectra.  These parameters, together with the rotation period, provide us with accurate self consistent values upon which to base ZDI maps.

ZDI was performed for the six stars studied so far, using a maximum entropy image reconstruction procedure, and representing the magnetic field as a combination of spherical harmonics (\cite[Donati et al. 2006]{Donati2006}).  The maps were based off of the computed LSD profiles, which was necessary to provide sufficient S/N.

\section{Early results}

\begin{figure}[bh]
\begin{center}
 \includegraphics[width=1.7in]{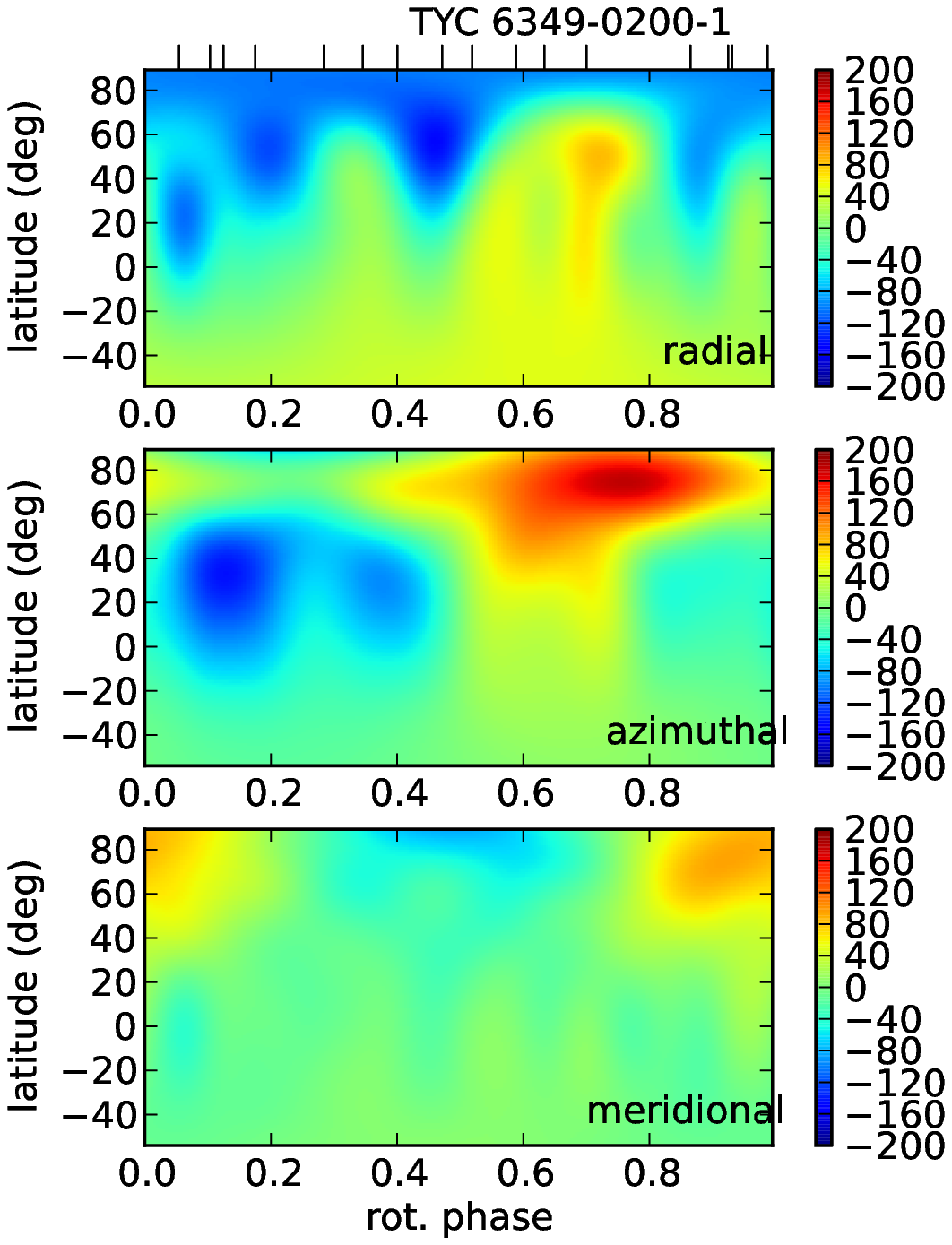} 
 \includegraphics[width=1.7in]{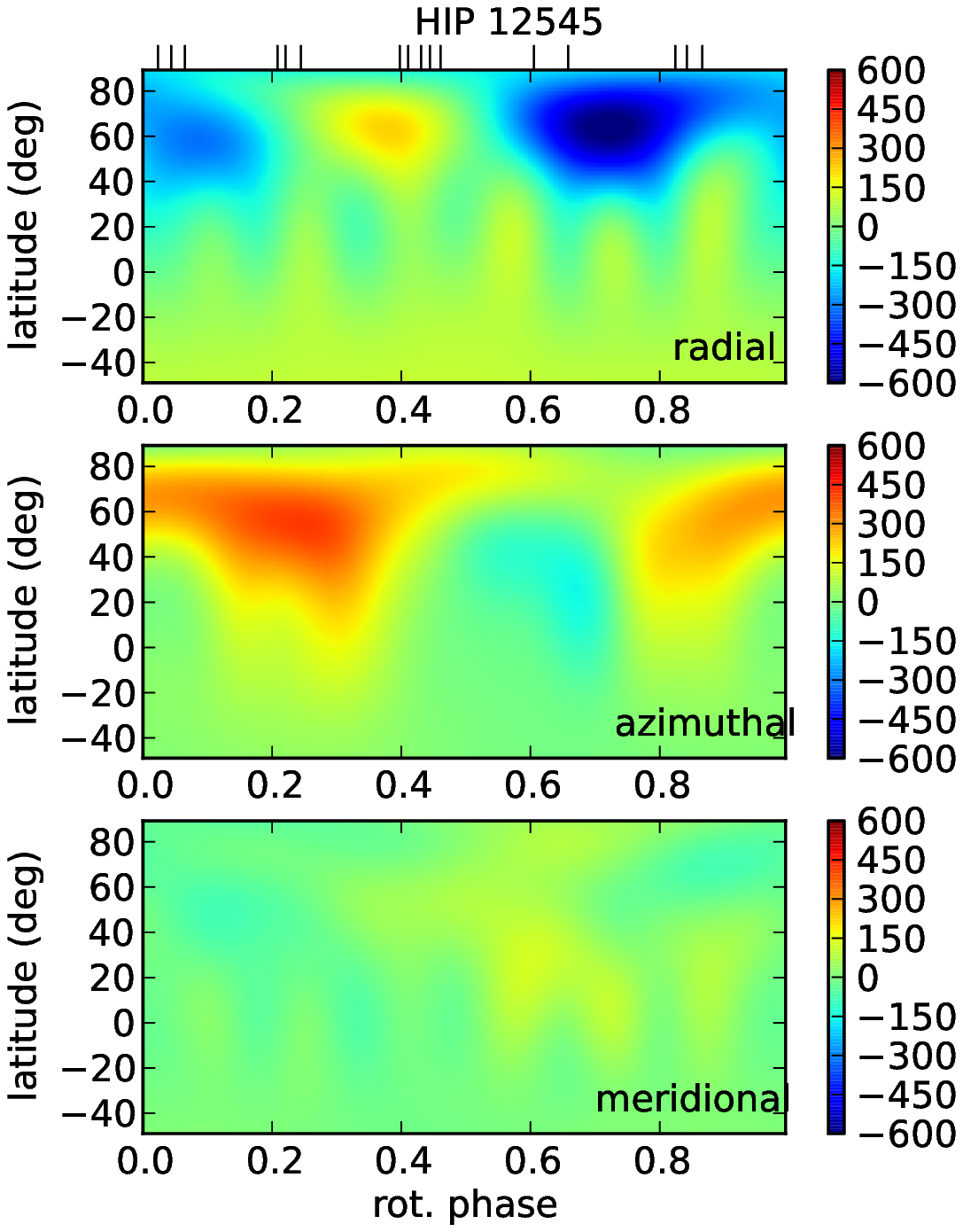} 
 \includegraphics[width=1.7in]{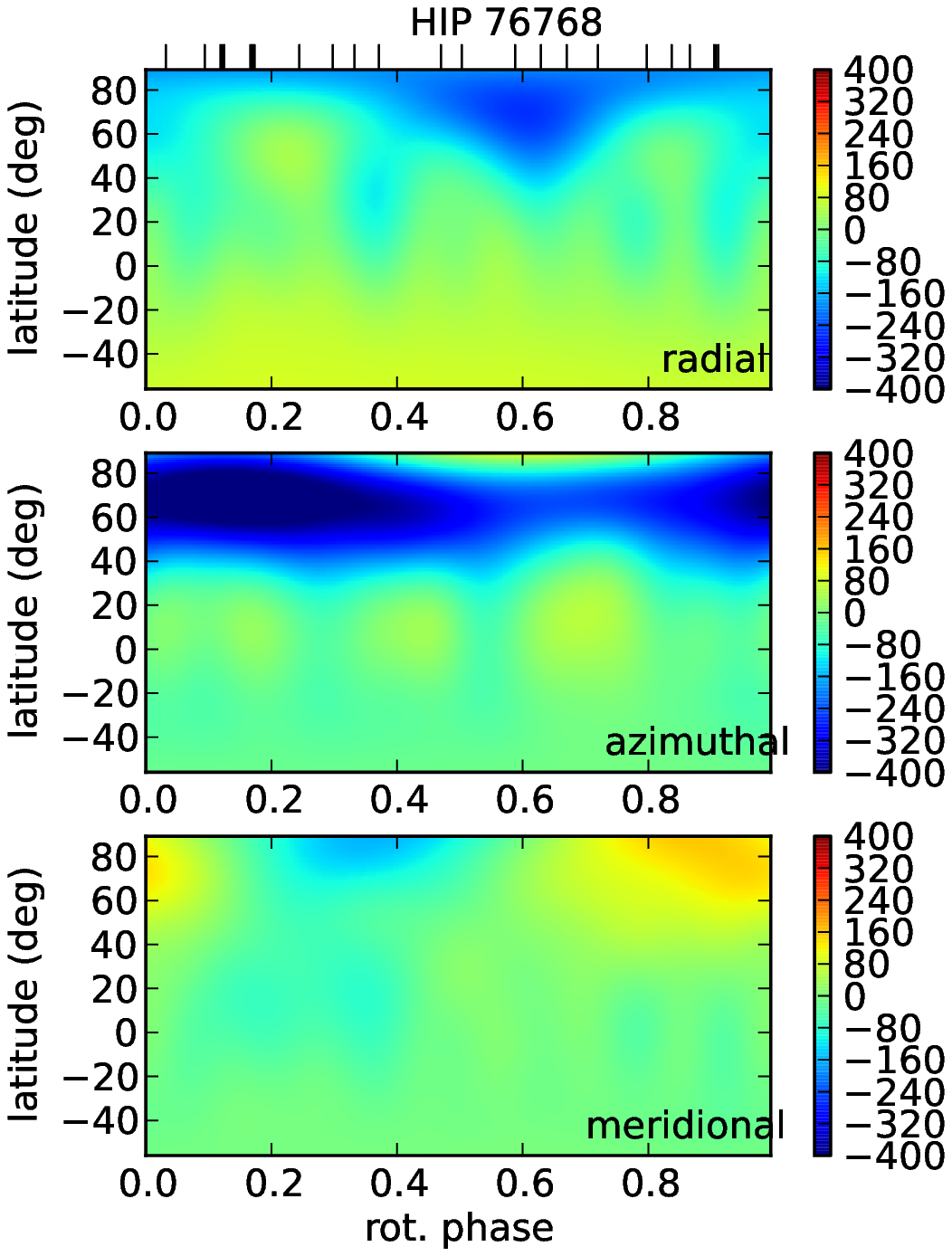} 
 \caption{Sample magnetic field maps, from ZDI, for the stars TYC~6349-0200-1, HIP~12545, and HIP~76768.  Plotted are the radial, azimuthal, and meridional components of the magnetic field.  Tick marks at the top of the figure indicate phases at which observations were obtained. }
   \label{fig1}
\end{center}
\end{figure}

Of the six stars analyzed so far, three are in the $\beta$ Pic association ($\sim$10 Myr): HIP~12545, TYC~6349-0200-1, and TYC~6878-0195-1, and three are in the AB Dor association ($\sim$100 Myr): HIP~76768, TYC~0486-4943-1, and TYC~5164-0567-1.  We find similar rotation periods for all six stars, from 3.4 days (TYC~6349-0200-1) to 5.7 days (TYC~6878-0195-1), with no clear distinction between the two associations.  We also find similar effective temperatures for the stars, from 4400 K (TYC~6349-0200-1) to 5100 K (TYC~5164-0567-1).  However, we find significant differences in the mean magnetic field modulus from the ZDI maps, from 40 G (TYC~0486-4943-1) to 180 G (HIP~12545).  We also find significant differences in the geometries of the reconstructed magnetic fields.  Three sample ZDI maps are presented in Fig. \ref{fig1}.

Trends in the evolution of the magnetic field with rotation period or age are weak at best for the small range of parameter space probed by these stars.  However, as this project progresses, it will be extended to associations with older ages, and to stars with a wider range of rotation periods.  This will enable us to draw much stronger conclusions than we can from the limited dataset currently available.

\end{document}